# Three-dimensional ultrasonic colloidal crystals

## *Cristaux colloïdaux ultrasonores tridimensionnels*

Mihai Caleap[*] and Bruce Drinkwater

*Faculty of Engineering, University of Bristol, BS8 1TR, United Kingdom*

## Abstract

Colloidal assembly represents a powerful method for the fabrication of functional materials. In this article, we describe how acoustic radiation forces can guide the assembly of colloidal particles into structures that serve as microscopic elements in novel acoustic metadevices or act as phononic crystals. Using a simple three-dimensional orthogonal system, we show that a diversity of colloidal structures with orthorhombic symmetry can be assembled with megahertz-frequency (MHz) standing pressure waves. These structures allow rapid tuning of acoustic properties and provide a new platform for dynamic metamaterial applications.

*L'assemblée colloïdale représente une méthode puissante pour la fabrication de matériaux fonctionnels. Dans cet article, nous décrivons comment les forces de rayonnement acoustique peuvent guider l'assemblage de particules colloïdales dans des structures qui servent comme éléments microscopiques dans les dispositifs à base de méta-matériaux acoustiques ou se comportent comme des cristaux phononiques. En utilisant un simple système orthogonal tridimensionnel, nous montrons qu'une diversité de structures colloïdales à symétrie orthorhombique peut être assemblée avec des ondes de pression stationnaires fonctionnant à des fréquences de l'ordre de quelques mégahertz (MHz). Ces structures permettent une tunabilité rapide des propriétés acoustiques et fournissent une nouvelle plate-forme pour les applications de méta-matériau dynamiques.*

## 1. Introduction

Photonic crystals have caused a paradigm-shift in the field of photonics, starting with the seminal work of Eli Yablonovitch[†] published in 1987[1] when these materials had never been manufactured, and were not known by that name. The concept of phononic crystals for sound waves followed a few years later. Photonic and phononic crystals share many common features, including the fact that they have both periodic internal structures. The effect of photonic crystals on light waves and phononic crystals on sound waves has been widely discussed. Today, photonic crystals appear in many areas of science, technology, medicine, and have also been discovered to be a product of nature in the biological world. There are also many potential applications of phononic crystals, and recent interest in these structures stems from the possibility of gaining previous unheralded control of sound waves for example, controlling the path of waves leads to incredibly efficient lenses[2] or invisibility cloaking[3,4] and controlling their transmission and reflection leads to highly efficient filters[5], diodes[6–8] or super absorbers[9].

---

[*] Corresponding author: Mihai.Caleap@bristol.ac.uk

[†] Given the plethora of applications, it is not surprising that he has recently received the Isaac Newton Medal from the Institute of Physics, for his contributions to the field of photonics.

In many situations, it is highly desirable to have the ability to tune the acoustic properties of such materials by physical means. Colloidal assembly represents a powerful bottom-up approach for the fabrication of functional materials, and simple three dimensional crystalline structures can be created by self-assembly, within the physical and thermo-dynamical confines of the system. Here, a colloidal crystal[10] stands for an ordered array of monodisperse colloidal particles, similar to a crystal in which repeating unit cells are atoms or molecules. In acoustics, colloidal crystals have established applications as phononic crystals[11–13], and this is the focus of this paper. Bulk properties of colloidal crystals depend on composition, particle size, packing arrangement, and degree of regularity. While the particle shape, size, volume fraction, charge, solvent screening lengths, *etc.* are important control parameters for colloidal phononic crystal experiments, none of them affords active control.

Acoustic radiation forces provide a convenient and effective tool for the assembly of colloidal particles into periodically arranged crystals and open up the possibility of tuning the diffraction of phononic structures. Active tuneability can be achieved by tailoring the lattice of the crystal, providing new opportunities for fundamental as well for applied research. Here, we present an overview of how colloidal structures can be assembled using acoustic radiation forces, and how the phononic properties of the resulting structures can be reconfigured and reversibly tuned by manipulating the acoustic force fields.

The organisation of the paper is as follows. We start by explaining how acoustic radiation forces can be used to trap and manipulate spherical microparticles in a fluid with megahertz-frequency (MHz) ultrasound waves. Here, we consider the case of standing and quasi-standing pressure waves created by a pair of transducers or a single transducer positioned opposite to a reflector. Then, combining two or three pairs of opposing transducers we discuss the diversity of colloidal structures that can be assembled using these as orthogonal systems. This results in a fully controllable colloidal crystal made of particles trapped in a reconfigurable pattern. The acoustophoretic motion of an agglomerate of fluid or solid particles suspended in aqueous solution is discussed; this closely mimics experimental particle tracing and velocimetry. Based on knowledge of the particle trajectories, an estimate of the theoretical reconfiguration time is presented.

## 2. Acoustic assembly of colloidal crystals

We have recently reported on the realization of an acoustic metadevice[14] that can generate three dimensional colloidal crystals, and demonstrated, for the first time, that it is possible to dynamically alter the geometry of the resulting crystal in real time. The reconfigurable colloidal crystal is assembled from microspheres in aqueous solution, trapped with acoustic radiation forces. The acoustic radiation force is governed by an energy landscape, determined by an applied high amplitude acoustic standing wave field, in which elastic particles move swiftly to energy minima. This creates a colloidal crystal of several cubic millilitres in volume with spheres arranged in an orthorhombic lattice in which the acoustic wavelength is used to control the lattice spacing.

The acoustic metadevice consists of two pairs of opposing parallel transducers, along with a single transducer, positioned on the base of the device in order to hold particles against gravity, Fig. 1. The levitation system, arranged vertically, lifts the particles and holds them in horizontal planes, while the manipulation system, uses counter-propagating waves to trap the particles in a grid of nodal positions. The generated acoustic landscape exerts an acoustic radiation force on the particles. The force is a second order nonlinear effect and stems from a combination of the time averaged pressure and inertial interaction between the particles and the acoustic field. It is this force that allows trapping of suspended particles at nodes or antinodes of that wave, depending on their radius as well as the compressibility and mass density of the particles and the host fluid. In the following, we expand on this narrative, and show the diversity of colloidal structures that can be assembled using a three dimensional orthogonal system.

### a. Acoustic particle trapping

Consider a spherical particle of size $a$, density $\rho$, and bulk modulus $\kappa$ in an inviscid fluid of density $\rho_0$ and bulk modulus $\kappa_0$. The particle is first assumed to be placed in a one dimensional (1D) standing wave field. Experimentally, a standing wave is created by counter-propagating waves from a pair of opposing transducers. Alternatively, the excitation of a single transducer positioned opposite to a reflector also can result in a standing wave. Here, our main concern is the acoustic radiation force acting on the particle.

#### 1. Standing pressure wave

Let us consider first the case of a standing pressure wave where two opposing travelling waves $\exp(\mathrm{i}(k_0 x + \Phi^+))$ and $\exp(-\mathrm{i}(k_0 x - \Phi^-))$ are excited simultaneously by a pair of transducers with different phase delays and $\Phi^+$ and $\Phi^-$, such as those in the acoustic manipulator[14] of Fig. 1. Here, $k_0 = 2\pi/\lambda$ is the wavenumber, $\lambda$ is the wavelength of acoustic wave in the fluid matrix, and $\omega = 2\pi f$ is the angular frequency. Assuming the opposing transducers generate identical sinusoidal pressure of amplitude $p_0$ with phase difference $\Delta\Phi$ $(= \Phi^+ - \Phi^-)$ between the two excitations, the resulting acoustic radiation force acting on a compressible sphere is then given by[15]

$$F = |p_0|^2 \frac{\pi}{\rho_0 \omega^2} \sin(2k_0 d - \Delta\Phi) \sum_{n=0}^{N \to \infty} (-1)^n (n+1)[(1 + 2S_n')S_{n+1}'' - (1 + 2S_{n+1}')S_n''], \quad (1)$$

where $d$ is the distance from the centre of the particle to the nearest pressure antinode. Note that any reflections at the front surface of the transducers are assumed to be small enough to be neglected. The coefficients $S_n = S_n' + iS_n''$ denote the multipolar scattering amplitudes for a spherical particle; they are determined by applying appropriate boundary conditions on the surface of the spherical particle. The radiation force vanishes if the sphere is centred on either a pressure antinode ($\Delta\Phi/2k_0 = 0$) or pressure node ($\Delta\Phi/2k_0 = \pm\pi/2$). The magnitude of the force is maximal when the particle is at the intermediate location ($\Delta\Phi/2k_0 = \pm\pi/4$). It is clear from Eq. (1) that depending on its location relative to the standing pressure wave, a particle can be forced either to the pressure node or pressure antinode. This is best demonstrated by considering the Rayleigh-scattering condition for which $a \ll \lambda$. It can be shown that the monopole ($n = 0$) and dipole ($n = 1$) terms are dominant in this limit. Let $\tilde{\omega} = k_0 a$ be the dimensionless frequency scattering parameter; then, considering only the leading terms in $\tilde{\omega}$, one has $S_n' = -S_n''^2$, ($n = 0,1$) with $S_0'' = \tilde{\omega}^3[\mathcal{M}/3 + \mathcal{O}(\tilde{\omega}^2)]$ and $S_1'' = \tilde{\omega}^3[\mathcal{D}/3 + \mathcal{O}(\tilde{\omega}^2)]$, where

$$\mathcal{D} = \frac{\rho - \rho_0}{\rho_0 + 2\rho} \quad \text{and} \quad \mathcal{M} = \frac{\kappa_0}{\kappa} - 1. \quad (2)$$

Then, the acoustic radiation force reduces to the well-known Gor'kov result[16], *i.e.*

$$F = |p_0|^2 \frac{\pi}{\rho_0 \omega^2} A\tilde{\omega}^3 \sin(2k_0 d - \Delta\Phi), \quad \tilde{\omega} \ll 1, \quad (3)$$

where $A = \mathcal{D} - \mathcal{M}/3$ is often referred to as the acoustopthoretic contrast factor. For dense particles and particles less compressible than the host fluid, such as polystyrene particles in aqueous solution, $A > 0$ and $F \propto +\sin(2k_0 d)$. For particles less dense and more compressible than the suspending fluid, such as some liquid droplets or air bubbles in water, we have $A < 0$ and $F \propto -\sin(2k_0 d)$. Consequently, the resulting particle motion is towards and away from the nodal plane, respectively. A sketch of the acoustic radiation force is shown in Fig. 2. Although Gor'kov formula (3) seems very practical for force calculation on small spheres, it is unfortunately not suitable for resonant particles and/or particles having arbitrary dimension. Conversely, expression (1) allows the force calculation to be effective for any frequency and wavelength. Figure 3a/b illustrates the acoustic radiation force exerted on a 90 $\mu$m-diamerer polystyrene/dodecane particle in aqueous solution.[‡] Both the exact formula (1) and Gor'kov result (3) are compared *versus* the frequency, for a standing wave with peak pressure amplitude $p_0 = 300\ kPa$. Here, and hereafter, the location $d$ of the particle is such that $\sin(2k_0 d - \Delta\Phi) = 1$. (*i.e.* $\Delta\Phi/2k_0 = \pi/4$ ). It is clear from Fig. 3 that the exact acoustic radiation force (1) exhibits a more irregular behaviour as the exciting frequency $f$ of the standing wave increases. When the particle size becomes comparable to the driving wavelength, the acoustic radiation force unveils intense peaks associated with waves guided by the surface of the particle which are weakly damped. These peaks are related to the resonance frequencies of an isolated particle. Here we observe that the solid particle is more resonant than the liquid particle. The resonance frequencies can be defined from the maximum magnitude of the multipolar scattering coefficients $|S_n|$, or from the frequency variations of their phase angle. As an example, Fig. 4 displays the scattering coefficients for a polystyrene particle in water. Note from Fig. 3a (3b) that when $F < 0$ ($F > 0$), although the acoustophoretic contrast factor $A > 0$ ($A < 0$), a polystyrene sphere (dodecane droplet) will be forced to a pressure antinode (node).

---

[‡] Acoustic property values for these materials are $\rho_0 = 0.997$ kg/m$^3$, $\kappa_0 = 2.2$ GPa for water; $\rho = 0.75$ kg/m$^3$, $\kappa = 1.23$ GPa, for dodecane; and $\rho = 1.06$ kg/m$^3$, $\kappa = 4.4$ GPa for polystyrene.

## 2. Quasi-standing pressure wave

Let us now consider the case of a partial standing wave such as that created in the acoustic levitator[14] of Fig. 1. A travelling wave generated by a single transducer with sinusoidal pressure of amplitude $p_0$ is reflected back by an opposite reflector with amplitude $p_0 R$. Assuming that secondary reflections can be ignored, the acoustic radiation force is obtained as a linear superposition of two forces, one resulting from a plane standing wave with acoustic energy density $|p_0|^2 R/(4\rho_0 c_0^2)$ and the other from a plane travelling wave with acoustic energy density $|p_0|^2(1 - R^2)/(4\rho_0 c_0^2)$; here, $c_0$ is the speed of sound in the host fluid. We obtain $F = RF_{st} + (R^2 - 1)F_{tr}$, where $F_{st}$ is the radiation force given by Eq. (1) and $F_{tr}$ is the radiation force acting on a particle due to a plane travelling wave,

$$F_{tr} = |p_0|^2 \frac{\pi}{\rho_0 \omega^2} \sum_{n=0}^{N\to\infty} (n+1)[S'_n + S'_{n+1} + 2(S'_n S'_{n+1} + S''_n S''_{n+1})]. \qquad (4)$$

Figure 5 illustrates the acoustic radiation force acting on a polystyrene particle due to a partial standing wave with two different reflection coefficients $R$, for the same parameters used in Fig. 3a. The fraction of momentum carried by a travelling wave is determined by the momentum carried away by the scattered wave, whereas for a standing wave there are important contributions from the interference between the excited field and scattered waves. The acoustic radiation force in the former case ($R = 0$) is always positive as seen in Fig. 5b, and the particle will be pushed in the direction of the wave propagation.

Note that the acoustic levitator in our metadevice is simply based on the reflections from the free surface between the aqueous solution and air. In practice $|p_0| > |p_0 R|$ due to propagation losses and beam divergence, although the reflection coefficient is approximately $-1$. If the water height inside the levitator is not exactly an integer multiple of $\lambda_z/2$, then a pressure gradient across the water/air surface will exist, giving rise to a force which will displace the water/air surface locally. If this displacement does not exceed limits imposed by surface tension, the water/air surface will shift until the equilibrium condition $p(H + \delta z) = 0$ is fulfilled ($p$ is the total acoustic pressure and $\delta z$ is an incremental shift in the water height) and the gradient across the surface is minimized; here the water height is adjusted to define a pressure antinode at $z = H$. Careful adjustment of the water height to obtain standing waves was expected in our experiment[17], but instead it was found that standing waves occurred regardless of the water height. Observe from Fig. 5 that the acoustic radiation force is negative in the low frequency range. For a reflection coefficient $R \simeq -1$, as is expected in the acoustic levitator, the radiation force is precisely the opposite of that shown in Fig. 3a. This may seem confusing at first, however the polystyrene particle will still be pushed to a pressure node. This is due to the fact that the position of the stable nodes changes as $\cos k_0 d$ changes in the case of a standing wave to $\sin k_0 d$ in the case of a quasi-standing wave.

b. Acoustic particle manipulation

Our acoustic metadevice, combines two pairs of opposing transducers in the $X$-$Y$ plane with the acoustic levitator operative in the third dimension ($Z$-axis), creating a fully controllable colloidal crystal made of particles trapped in a three dimensional reconfigurable pattern. One transducer and a parallel reflector, such as the acoustic levitator in our device, will result in a pattern with $\pi$ rotational symmetry and a translational symmetry in one dimension with a period of $\lambda_z$, and linear nodes separated by $\lambda_z/2$, as already inferred in the previous discussion. Two pairs of orthogonal transducers result in a pattern with $\pi/2$ symmetry and translational symmetry in two perpendicular directions. When the manipulation transducers are excited in pairs, the separation between the lines in the grid pattern are $\Delta x = \lambda_x/\sqrt{2}$ and $\Delta y = \lambda_y/\sqrt{2}$. It is possible to translate the entire grid pattern by applying further phase changes to the excitations. The spacing of the pattern depends on the frequency used and it becomes more densely spaced as the frequency increases. A sketch demonstrating the acoustic manipulation using a pair of transducers is shown in Fig. 6.

Assuming that there is negligible reflection from the transducer faces, the position of traps changes linearly with the relative phases, $\Delta\Phi_x$ and $\Delta\Phi_y$, between the excitation signals applied to each pair of transducers. In practice, multiple reflections between transducers lead to a deviation from the linear relationship between translation distance and relative phases; thus we would expect to see deterioration in performance. However, it has been shown that the reflection can become sufficiently low in the vicinity of through-thickness resonances of the piezoelectric transducers,[18] creating good conditions for controlled particle manipulation. As an example, Fig. 7 shows the measured impedance magnitude and phase characteristics for 2 mm and 3 mm-thick piezoelectric transducers, such as those used in our acoustic metadevice. The phase-jumps clearly indicate the resonance frequencies where $|Z|$ reaches maxima. By tuning the applied frequency to one of these resonance frequencies, the acoustic forces are able to reliably manipulate polystyrene particles suspended in the aqueous solution.

### c. Acoustic lattice reconfigurability

The reconfiguration speed is a key point for reconfigurable metadevice applications; here, we estimate the theoretical reconfiguration time and its feasible limit, based on knowledge of the particle trajectories.

The time taken for particles to move between different patterns is governed by a combination of the acoustic forces on the particles and viscous drag. Consider the case of a simple 1D standing wave, as generated by a pair of transducers in the acoustic manipulator of our device. The trajectory of a single small ($\widetilde{\omega} \ll 1$) spherical particle is given by $k_0 x = \arctan\left[\exp\left(A \frac{(p_0 \widetilde{\omega})^2}{9\eta \rho_0 c_0^2} t\right) \tan k x_{in}\right]$,[19] where $\eta$ is the dynamic viscosity of the suspending fluid, and $x_{in}$ is the initial position. If we assume that $\eta \approx 0.9\, k\text{Pa} \cdot \text{s}$ and take $x_{in} = 0.1\lambda/4$ and $x_f = 0.99\lambda/4$, then, the time $t$ to effect the axial movement of a $90\, \mu m$-diameter polystyrene particle at, say 5.25 MHz, is $\sim 5\, ms$ for the conditions of interest here. The pressure amplitude was taken as $p_0 = 300\, k\text{Pa}$; the upper pressure would be limited in practice by cavitation to the order of 1 MPa, leading to a minimum response time of $< 0.8\, ms$, assuming all other parameters remain constant. This in turn, determines the switching, or reconfiguration time for the metadevice.

### d. Particle control in two dimensions

In order to elucidate how a trap can be created in two dimensions (2D), consider two pairs of opposing transducers, such as those of the manipulation stage, in the *X-Y* plane. Omitting the harmonic time dependence, the total pressure field can be written as a sum of the contributions from each transducer, as

$$p = 2p_0\left[e^{i\phi_x} \cos(k_0 x + \Delta\Phi_x/2) + e^{i\phi_y} \cos(k_0 y + \Delta\Phi_y/2)\right], \tag{5}$$

where $\phi_j = (\Phi_j^+ + \Phi_j^-)/2$, $\Delta\Phi_j = \Phi_j^+ - \Phi_j^-$; here, $\Phi_j^\pm$ (with $j = x, y$) denote the phase delays applied to each excitation.

Rather than having two pairs of opposing transducers, a single pair together with a reflector opposing another transducer may also be used to generate two dimensional standing waves. This condition is met in our acoustic metadevice if one pair of transducers in the manipulation stage is switched off and the levitation stage is on, *e.g.* in the *X-Z* plane. Assuming that the reflector is the water/air interface of reflection coefficient $R \simeq -1$, the total pressure field can be written as

$$p = 2p_0\left[e^{i\phi_x} \cos(k_0 x + \Delta\Phi_x/2) + i e^{i\Phi_z^\pm/2} \sin(k_0 z + \Phi_z^\pm/2)\right]. \tag{6}$$

Figure 8 illustrates the absolute pressure field resulting from the superposition of two orthogonally oriented cosine functions [Eq. (5)], or a sine and a cosine function [Eq. (6)], with identical amplitude, frequency and phase. These results are displayed for a single wavelength $\lambda$ in both $X$ and $Y$ (or $Z$) directions. The force field is indicated by arrows, and is calculated using the simple Gor'kov formula (3) extended to 2D planar standing waves. The length of the arrows is proportional to the magnitude of the acoustic radiation force, while the orientation of the arrow indicates the direction of the force. Observe that particles collect at two or three locations per wavelength, corresponding to the minima or maxima of the pressure field; these locations are indicated by solid green dots. Small elastic particles with positive acoustophoretic factor ($A > 0$), *e.g.* polystyrene spheres, tend to pressure nodes (Fig. 8a and c), while, particles with $A < 0$, *e.g.* dodecane droplets, tend to pressure antinodes (Fig. 8b and d). Note that results similar to Fig. 8c and d could also be obtained by adjusting the phases of one of the two opposed pairs, *i.e.* a sine wave is just a phase shifted cosine wave.

In order to study the acoustophoretic motion of an agglomerate of particles suspended in aqueous solution, we apply the COMSOL Particle Tracing Module which closely mimics experimental particle tracing and velocimetry. This model is that of the geometry of Fig. 1 driven at a constant frequency, with plane wave radiation conditions at the open boundaries and matched boundary conditions at the water/transducer interface. Elastic particles (with $A > 0$) of different sizes are released uniformly in the standing pressure field and their path is determined when influenced by the acoustic radiation force, viscous drag, and gravity. In this Module, the implemented acoustic radiation force reduces exactly to the analytical result of Eq. (3) for (1D) planar standing waves.

The pressure field amplitude at 1 MHz, with opposing transducers excited in-phase ($\phi_x = \phi_y$), or out-of-phase ($\phi_x = \phi_y + \pi/2$), is shown in Fig. 9. Figure 10 depicts the positions of small, medium, and large elastic particles after $t = 1$ s. Depending on their size, particles form wire-grid patterns, or square patterns. Observe that when the transducers are out-of-phase, a grid of localised traps (or regions of zero pressure) where particles are locked in a square pattern, is more rapidly and uniformly created than with all transducers in phase. Although the results of Fig. 10 were specifically obtained for polystyrene particles in water, with diameters of 10 $\mu$m (small), 50 $\mu$m (medium), and 90 $\mu$m (large), we have verified that similar behaviour is observed for a variety of materials (for which $A > 0$). Similar arguments apply to particles with $A < 0$ (*e.g.* dodecane droplets), with patterns resembling those of Fig. 10d-e-f.

In order to give a qualitative comparison with the results of Fig. 10, we present next experimental micrographs of assembled colloidal crystals obtained using the acoustic metadevice[17] of Fig. 1. The experiment was performed using 90-μm-diameter polystyrene spheres in aqueous suspension contained within a central cylindrical chamber 10 mm in diameter. Each of the piezoceramic transducers (3 mm-thick PZT plates) was driven using a continuous sine wave by a signal generator. Each transducer pair was synchronized to allow the relative phase of the excitations of the opposing transducers to be adjusted. Synchronization between channels was achieved using two arbitrary waveform generators providing four outputs allowing independent control of the amplitude, phase, and frequency of the transducer pairs. The four manipulation transducers are excited at $f = 0.725$ MHz

(corresponding to the fundamental thickness-mode resonance) and the excitation voltage is 20 $V_{p-p}$. Figure 11 shows views of the *X-Y* plane of the colloidal crystal assembled with pairs of opposing transducers excited in phase (with $\Delta\Phi_x = \Delta\Phi_y = 0$, $\phi_x = \phi_y = 0$) or out-of-phase with $\Delta\Phi_x = \Delta\Phi_y = 0$, $\phi_x = 0$, $\phi_y = \pi/2$). It is clear from Fig. 11 that the position of particles resembles that shown in Fig. 10. As expected, the particles move toward the pressure nodes for the frequency considered.

e. **Particle control in three dimensions**

Introducing an additional standing wave enables more complex acoustic landscapes to be produced. Similar to Fig. 9, the result in Fig. 12 illustrates the force field exerted by a three dimensional standing pressure wave on small particles in aqueous solution. It is clear that elastic particles tend to pressure nodes (Fig. 12a and c) creating a simple cubic (sc) crystal, while fluid particles tend to pressure antinodes (Fig. 12b and d) creating a body-centred cubic (bcc) crystal.

Note however, that the results of Fig. 12 are obtained from the superposition of three orthogonally oriented cosine functions, or a sine and two cosine functions, with identical amplitude, frequency and phase. In principle, for different operating wavelengths $\lambda_x$, $\lambda_y$, and $\lambda_z$, more elaborate patterns can be created, with general orthorhombic symmetry.

In addition, a combination of fluid and solid particles would create a binary colloidal super-lattice crystal, for example, with elastic particles trapped at the antinodes and fluid particles at the nodes. These colloidal crystals present unique opportunities to create novel interconnected microstructures. The addition of more transducers would provide opportunities for the creation of a wider variety of colloidal structures. Indeed, multi-element array devices[20] would further extend the versatility, and could actively control the detail of the lattice geometry. Fabricating perfect colloidal crystals is in practice quite challenging, since small imperfections in the symmetry of lattice could always occur during the assembly process, especially when dealing with large volumes of colloidal suspensions. Intrinsic defects such as missing or multiple particles at some nodes (see Fig. 11) disrupt the perfect periodicity of the crystalline lattice and create specific acoustic propagation behaviours within the band gaps. However, in photonic crystals it has been shown that various interesting behaviours occur around defects[21,22], *e.g.* the ability to trap energy within localised defect regions. Energy localisation around defects in both phononic and photonic materials is an on-going and active area of research.[23]

**Conclusion**

It has been demonstrated that a variety of ultrasonic colloidal crystals can be created with standing pressure waves, in the megahertz range. This opens up opportunities both in colloidal science and in the rapidly evolving fields of phononic crystals and acoustic metamaterials. The methodology described here will contribute to the search for the design of optimal metamaterial structures operating not only in the ultrasonic regime, but also for electromagnetic waves, in the few hundreds of terahertz range. This technique can also provide a platform for performing bio-assays and cell-assays in a non-contact way using ultrasonic standing waves.

**Figure list**

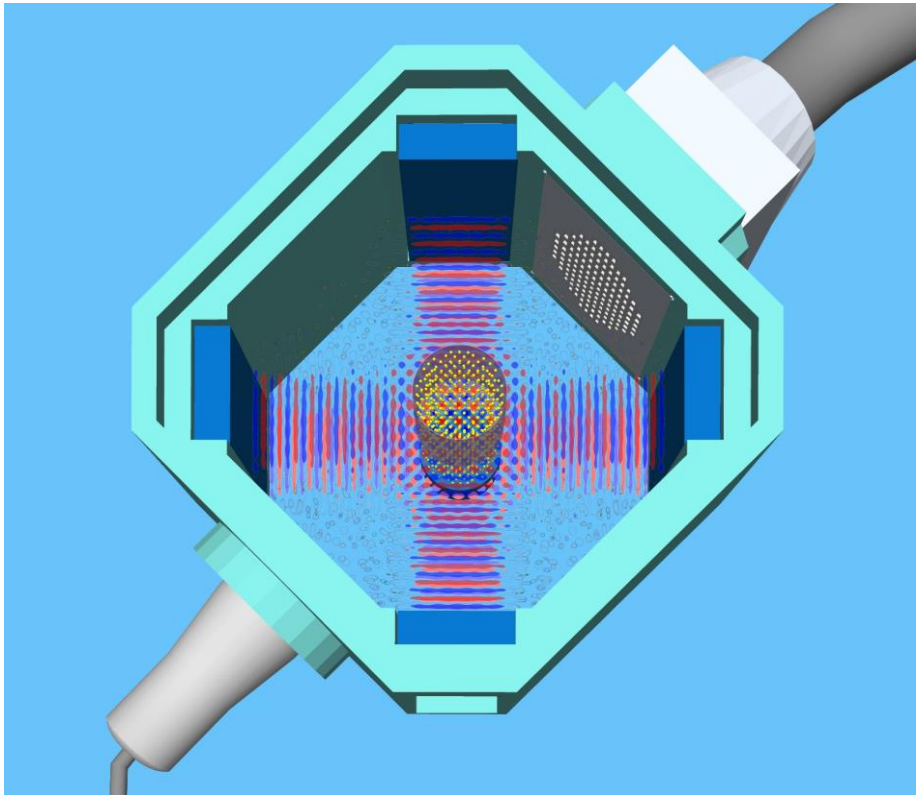

**Figure 1** Acoustic meta device with four transducers positioned as two orthogonal pairs, and one transducer at the base, which holds particles against gravity in horizontal planes inside a thin-wall transparent polyester tube.

*Dispositif acoustique avec quatre transducteurs placés comme deux paires orthogonales et un autre transducteur à la base, qui garde les particules contre la gravitation dans des plans horizontaux à l'intérieur d'un tube de polyester de mur mince transparent.*

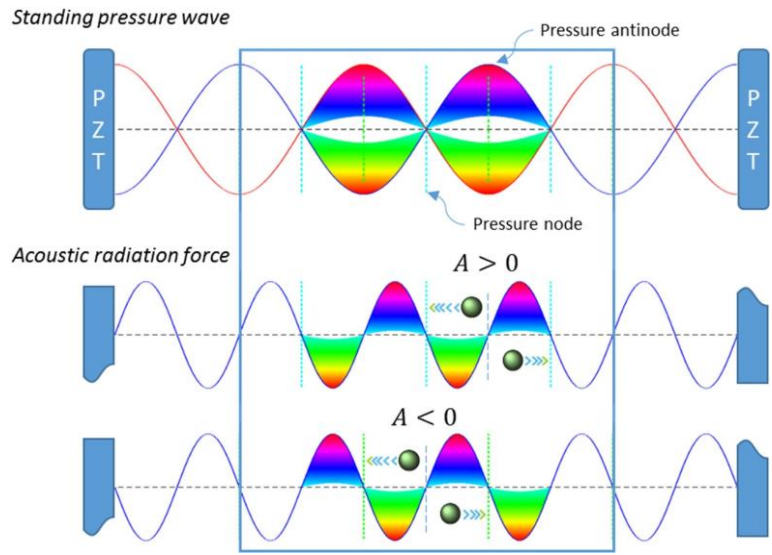

**Figure 2** Sketch of a standing pressure wave created by two parallel piezoceramic (PZT) plates. Relative to the pressure wave, the acoustic radiation force on a small suspended particle is period doubled and phase shifted. The acoustic radiation forces drive particles to nodes or antinodes in the standing pressure wave, depending on the sign of the acoustophoretic contrast factor $A$.

*Schéma d'une onde de pression stationnaire créée par deux plaques parallèles piézo-électriques (PZT). Par rapport à l'onde de pression, la force de rayonnement acoustique sur une petite particule en suspension est doublée en période et décalée en phase. Les forces de rayonnement acoustique guident les particules aux nœuds ou aux ventres de pression, selon le signe du facteur de contraste acoustophorétique $A$.*

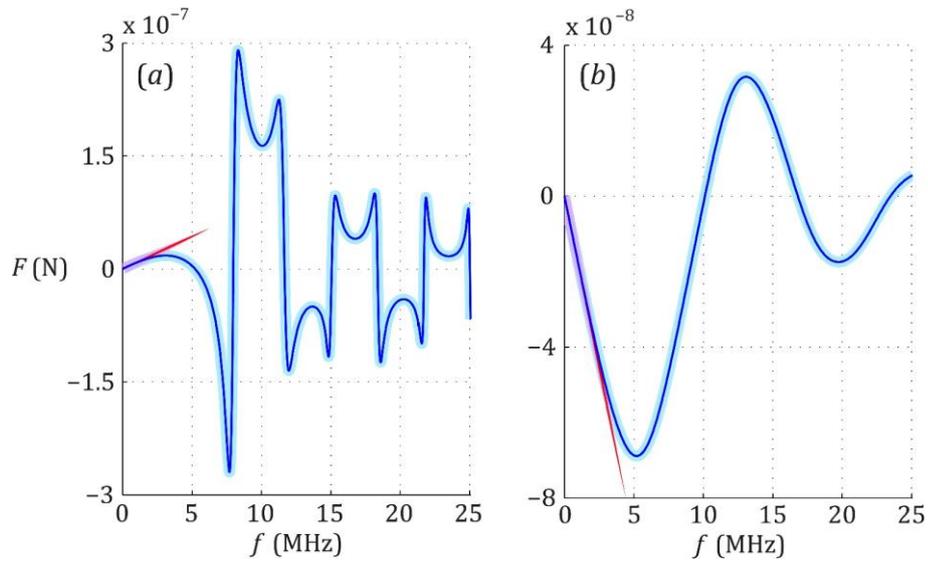

**Figure 3** Acoustic radiation force exerted by a standing wave, $p_0[\exp(ik_0 x) + \exp(-ik_0 x)]$, with pressure amplitude $p_0 = 300\ k\text{Pa}$, on a $90\ \mu$m-diameter polystyrene particle (a) or dodecane droplet (b) suspended in aqueous solution. A comparison with the Gor'kov result is depicted in the low frequency range. Here, the location of the particles is such that $k_0 d = 0$.

*Force de rayonnement acoustique exercée par une onde de pression stationnaire, $p_0[exp(ik_0 x) + exp(-ik_0 x)]$, d'amplitude $p_0 = 300\ kPa$, sur une particule en polystyrène (a) ou une goutte de dodécane (b) avec un diamètre de $90\ \mu m$, en suspension dans une solution aqueuse. Une comparaison avec le résultat de Gor'kov est représentée dans le régime de basses fréquences. Ici, la position des particules est telle que $k_0 d = 0$.*

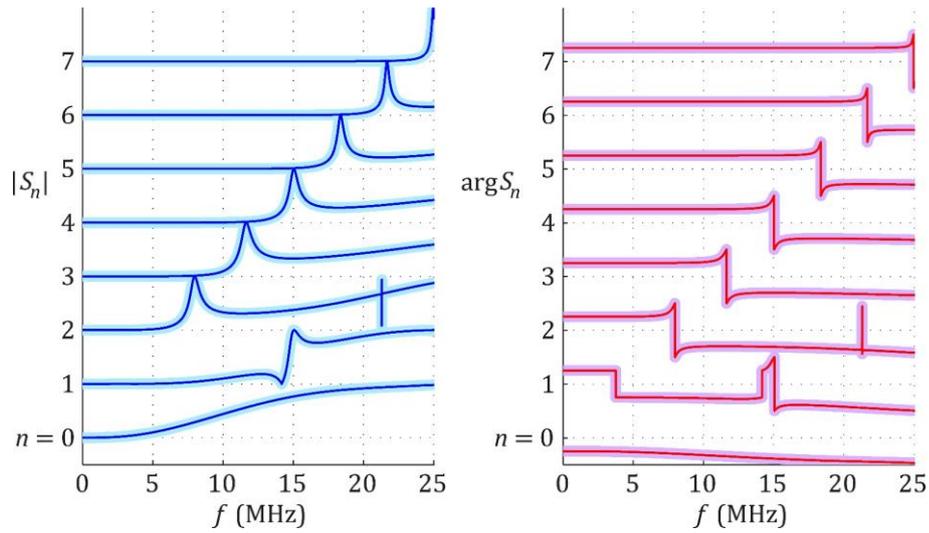

**Figure 4** Magnitude and phase of the multipolar scattering amplitudes $S_n$ *versus* the frequency $f$ for a 90 $\mu$m- diameter polystyrene particle suspended in aqueous solution. The phase-jumps (from $+\pi$ to $-\pi$) indicate the resonance frequencies where $|S_n|$ reaches maxima.

*L'amplitude et la phase des coefficients de diffusion multipolaires $S_n$ en fonction de la fréquence f pour une particule en polystyrène de 90 $\mu$m en diamètre, suspendue dans une solution aqueuse. Les sauts de phase (de $+\pi$ à $-\pi$) indiquent les fréquences de résonance où $|S_n|$ atteint de maxima.*

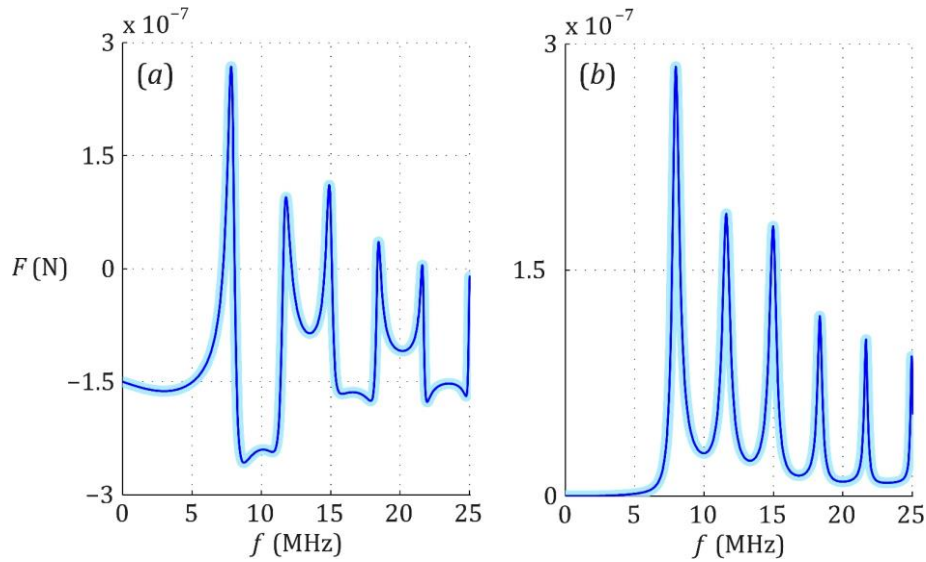

**Figure 5** Acoustic radiation force exerted by a partial standing wave, $p_0[\exp(ik_0 x) + R\exp(-ik_0 x)]$, with pressure amplitude $p_0 = 300\ k\text{Pa}$, on a 90 $\mu$m-diameter polystyrene particle suspended in aqueous solution. (a) $R = -0.5$ (quasi-standing wave); (b) $R = 0$ (plane travelling wave). Here, the location of the particles is such that $k_0 d = 0$.

*Force de rayonnement acoustique exercée par une onde de pression stationnaire partielle, $p_0[exp(ik_0 x) + Rexp(-ik_0 x)]$, d'amplitude $p_0 = 300\ k\text{Pa}$, sur une particule en polystyrène avec un diamètre de 90 $\mu$m, suspendue dans une solution aqueuse. (a) $R = -0.5$ (onde quasi-stationnaire) ; (b) $R = 0$ (onde plane progressive). Ici, la position des particules est telle que $k_0 d = 0$.*

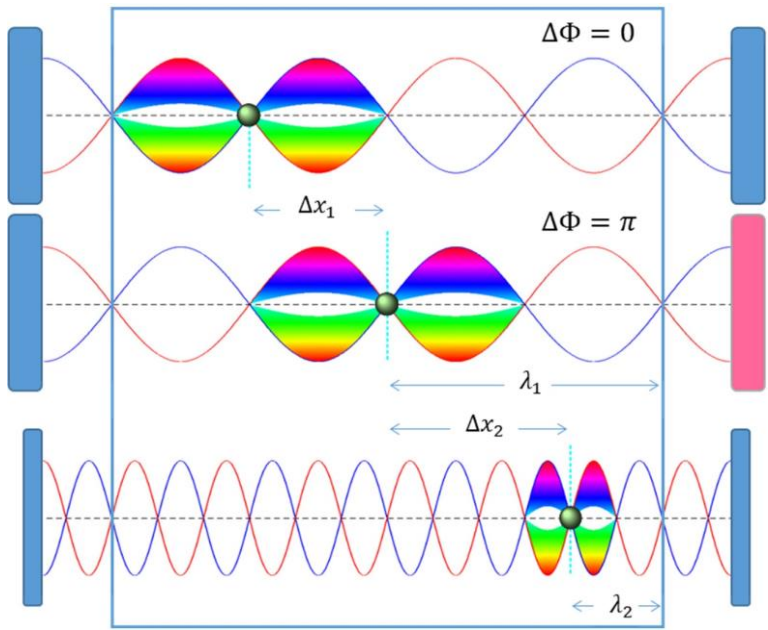

**Figure 6** Sketch of a standing pressure wave created by two parallel piezoceramic (PZT) plates. By varying the phase difference between the pair of waves or by using changes in frequency the nodal positions can be accurately controlled.

*Schéma d'une onde de pression stationnaire créée par deux plaques parallèles piézo-électriques (PZT). En faisant varier la différence de phase entre la paire d'ondes ou à l'aide de changements de fréquence, les positions nodales peuvent être contrôlées avec précision.*

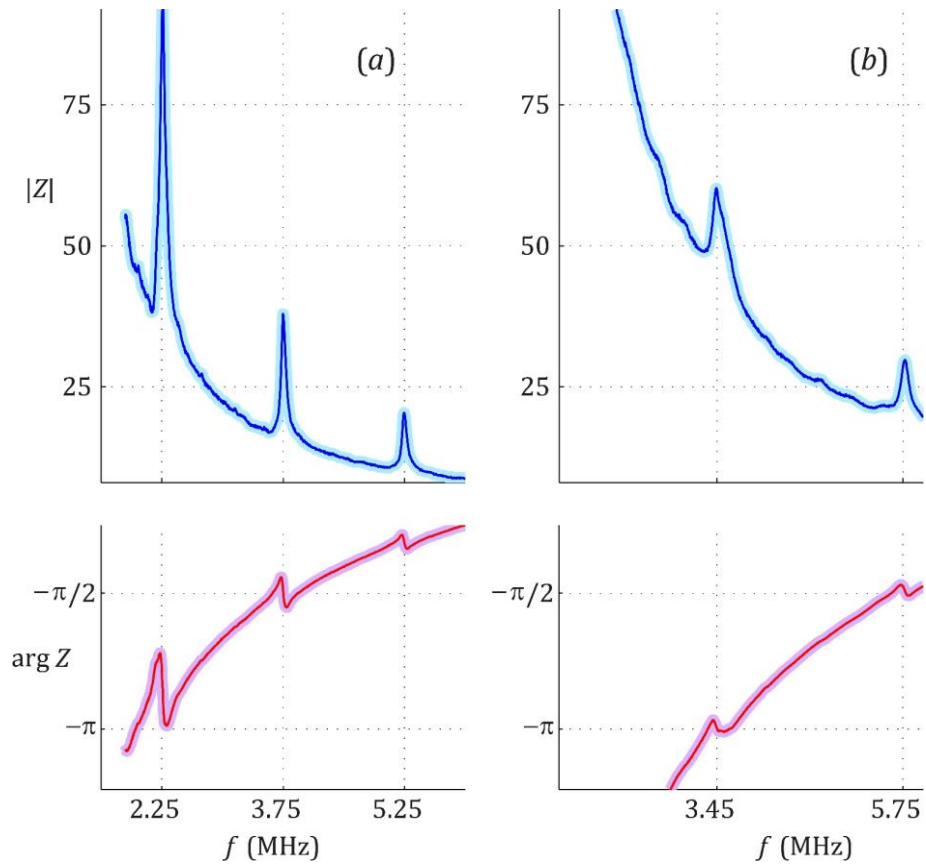

**Figure 7** Measured values for the impedance magnitude (in units of ohms) and phase for the manipulation (a) and levitation (b) transducers.

*Valeurs mesurées de l'amplitude (en unités d'ohms) et la phase de l'impédance pour les transducteurs de manipulation (a) et de lévitation (b).*

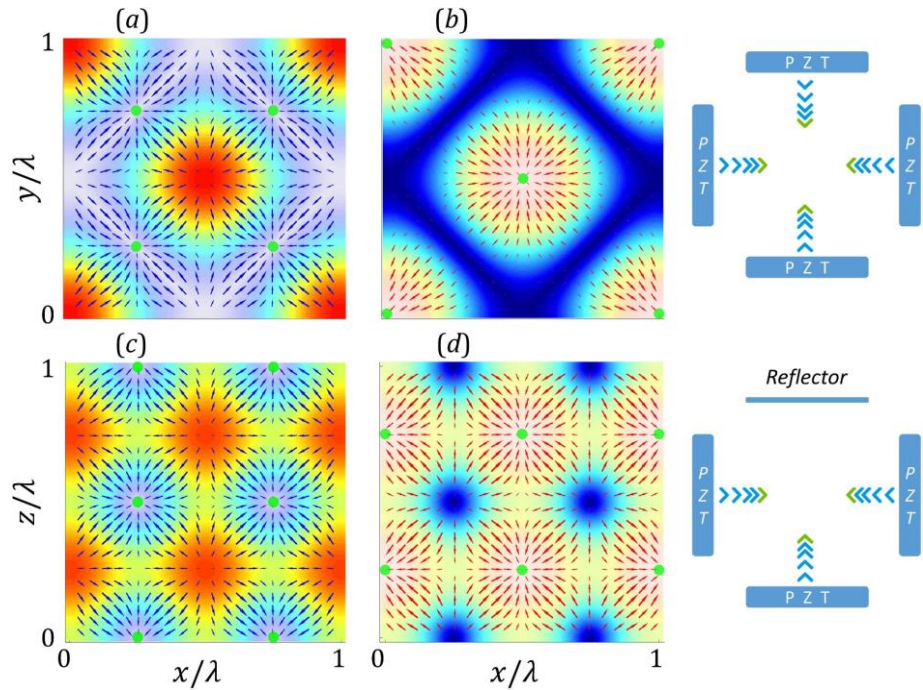

**Figure 8** Acoustic force field exerted by a two-dimensional standing pressure wave on spherical particles in aqueous solution. The length of the arrows is proportional to the magnitude of the force, while their orientation indicates the direction of the force. (a-c) Elastic particles; (b-d) Fluid particles. The sketch on the right illustrates the corresponding total pressure field, with $p = 2p_0(\cos k_0 x + \cos k_0 y)$ - top panels (a-b), and $p = 2p_0(\cos k_0 x + i \sin k_0 y)$ – bottom panels (c-d).

*Champ de force acoustique exercé par une onde de pression stationnaire à deux dimensions sur des particules sphériques en solution aqueuse. La longueur des flèches est proportionnelle à l'intensité de la force, tandis que leur orientation indique la direction de la force. (a-c) Particules élastiques ; (b-d) Particules fluides. Le schéma sur la droite illustre le champ de pression total correspondant, avec $p = 2p_0(\cos k_0 x + \cos k_0 y)$ – panneaux supérieurs (a-b) et $p = 2p_0(\cos k_0 x + i \sin k_0 y)$ – panneaux inférieurs (c-d).*

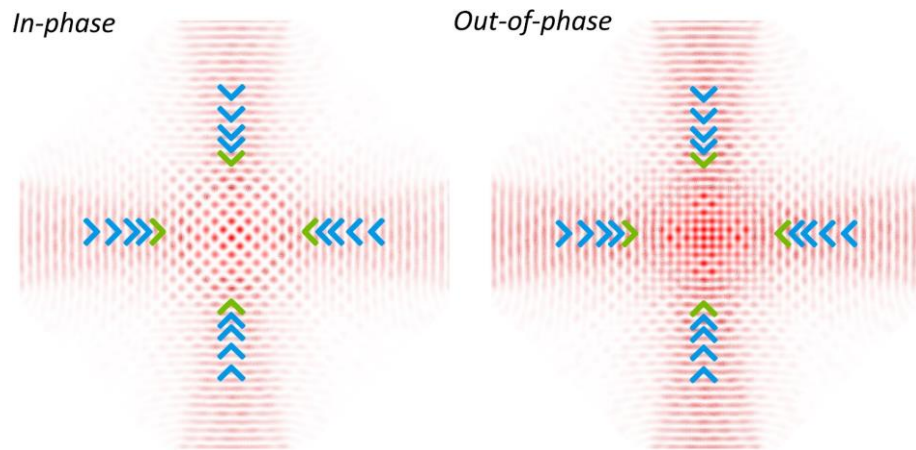

**Figure 9** Acoustic landscapes obtained with opposing transducers excited in-phase ($\phi_x = \phi_y$), or out-of-phase ($\phi_x = \phi_y + \pi/2$).

*Paysages acoustiques obtenus avec des transducteurs opposés excités en phase ($\phi_x = \phi_y$), ou en opposition de phase ($\phi_x = \phi_y + \pi/2$).*

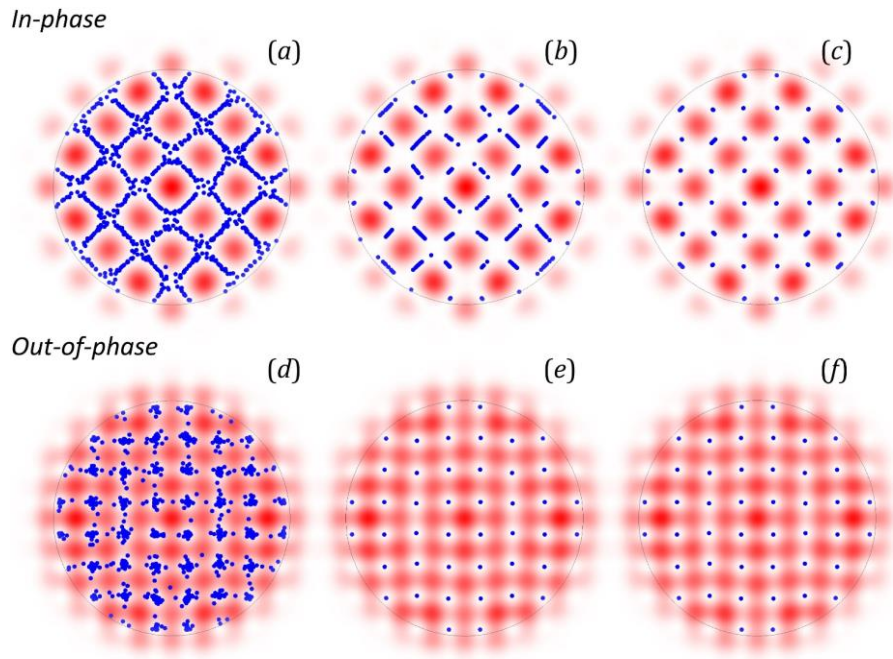

**Figure 10** Positions of small (a and d), medium (b and e), and large (c and f) elastic particles after $t = 1\ s$, with opposing transducers excited in-phase – top panels (a-b-c), or out-of-phase – bottom panels (d-e-f). This is an expanded view of the central circular regions of Fig. 9.

*Positions des particules élastiques petites (a et d), moyennes (b et e), et grosses (c et f) après $t = 1\ s$, avec des transducteurs opposés excités en phase – panneaux supérieures (a-b-c), ou en opposition de phase – panneaux inférieurs (d-e-f). Il s'agit d'une vision élargie des régions centrales circulaires de la Fig. 9.*

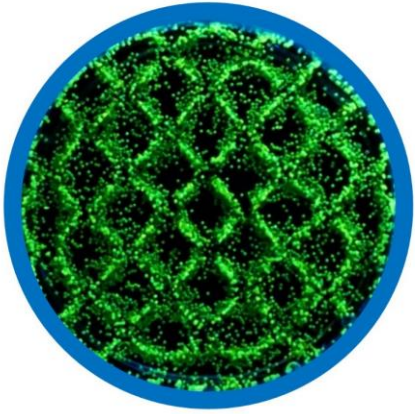 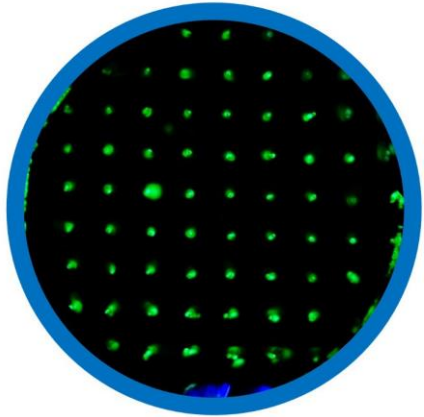

**Figure 11** Micrographs of the *X–Y* plane of the assembled colloidal crystals using the acoustic metadevice of Fig. 1. View from the top showing the positions of 90-μm–diameter polystyrene particles (after $t = 1\,s$) inside the central cylindrical region of 10-mm–diameter, with opposing transducers excited in-phase or out-of-phase.

*Micrographies du plan X-Y des cristaux colloïdaux, assemblés en utilisant le dispositif acoustique de la Fig. 1. Vue de haut, montrant les positions des particules de polystyrène avec un diamètre de $90\,\mu m$ (après $t = 1\,s$) à l'intérieur de la région centrale circulaire de 10 mm de diamètre, avec des transducteurs opposés excités en phase ou en opposition de phase.*

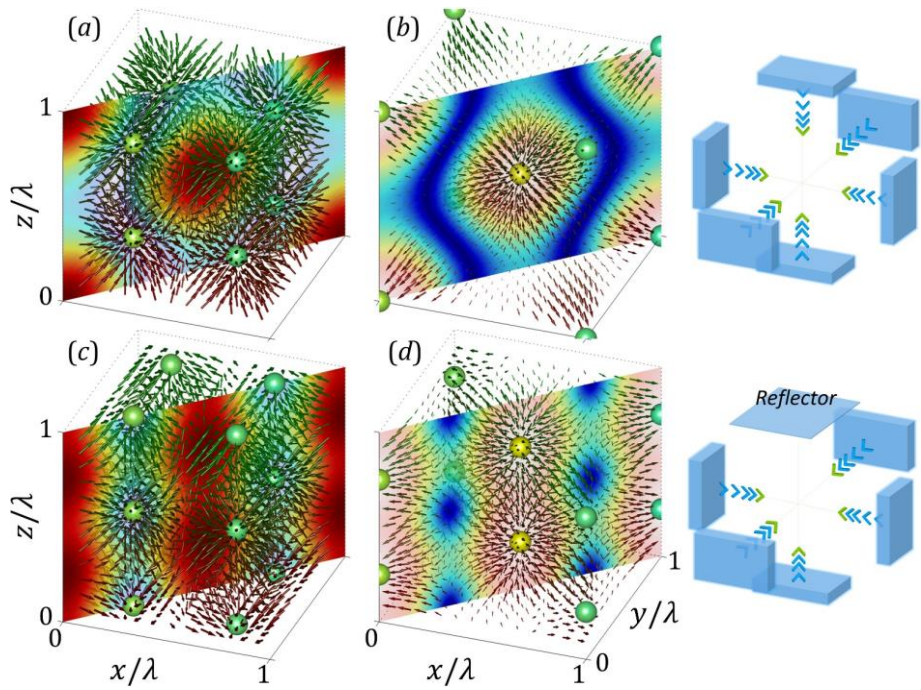

**Figure 12** Same as in Fig. 8, but in three dimensions, with $p = 2p_0(\cos k_0 x + \cos k_0 y + \cos k_0 z)$ - top panels (a-b), and $p = 2p_0(\cos k_0 x + \cos k_0 y + i \sin k_0 z)$ – bottom panels (c-d).

*Similaire à la Fig. 8, mais en trois dimensions, avec $p = 2p_0(\cos k_0 x + \cos k_0 y + \cos k_0 z)$ - panneaux supérieurs (a-b), et $p = 2p_0(\cos k_0 x + \cos k_0 y + i \sin k_0 z)$ – panneaux inférieurs (c-d).*